\documentclass[aps,prl,twocolumn,showpacs,lengthcheck,preprintnumbers]{revtex4}%
\usepackage{amsfonts}
\usepackage{amsmath}
\usepackage{amssymb}
\usepackage{graphicx}%
\begin{document}
\preprint{Phys. Rev. A \textbf{74}, 065602 (2006)}
\title{Extension of the Thomas-Fermi approximation for trapped Bose-Einstein
condensates with an arbitrary number of atoms}
\author{A. Mu\~{n}oz Mateo}
\email{ammateo@ull.es}
\author{V. Delgado}
\email{vdelgado@ull.es}
\affiliation{Departamento de F\'{\i}sica Fundamental II, Universidad de La Laguna, La
Laguna, Tenerife, Canary Islands, Spain}


\begin{abstract}
By incorporating the zero-point energy contribution we derive simple and
accurate extensions of the usual Thomas-Fermi (TF) expressions for the
ground-state properties of trapped Bose-Einstein condensates that remain valid
for an arbitrary number of atoms in the mean-field regime.\ Specifically, we
obtain approximate analytical expressions for the ground-state properties of
spherical, cigar-shaped, and disk-shaped condensates that reduce to the
correct analytical formulas in both the TF and the perturbative regimes, and
remain valid and accurate in between these two limiting cases. Mean-field
quasi-1D and -2D condensates appear as simple particular cases of our
formulation. The validity of our results is corroborated by an independent
numerical computation based on the 3D Gross-Pitaevskii equation.

\end{abstract}
\maketitle


The experimental realization of Bose-Einstein condensates (BECs) of dilute
atomic gases confined in optical and magnetic traps \cite{BEC1,BEC2,BEC3} has
stimulated great activity in the characterization of these quantum systems. Of
particular interest are the ground-state properties of trapped BECs with
repulsive interatomic interactions \cite{Baym1}. These properties derive from
the condensate wave function $\psi(\mathbf{r})$ which, in the zero-temperature
limit, satisfies the stationary Gross-Pitaevskii equation (GPE)
\cite{RevStrin}%

\begin{equation}
\left(  -\frac{\hbar^{2}}{2m}\nabla^{2}+V(\mathbf{r})+gN\left\vert
\psi\right\vert ^{2}\right)  \psi=\mu\psi, \label{TF0}%
\end{equation}
where $N$ is the number of atoms, $g=4\pi\hbar^{2}a/m$ is the interaction
strength, $a$ is the s-wave scattering length, $V(\mathbf{r})=\frac{1}%
{2}m(\omega_{\bot}^{2}r_{\bot}^{2}+\omega_{z}^{2}z^{2})$ is the harmonic
potential of the confining trap, and $\mu$ is the chemical potential.

Only in two limiting cases can Eq. (\ref{TF0}) be solved analytically: in the
Thomas-Fermi (TF) and perturbative regimes. When $N$ is sufficiently large
that $\mu\gg\hbar\omega_{\bot},\,\hbar\omega_{z},$ one enters the TF regime.
In this case the kinetic energy can be neglected in comparison with the
interaction energy and the GPE reduces to a simple algebraic equation. Useful
analytical expressions can then be obtained for the condensate ground-state
properties \cite{Baym1}. In the simple case of a spherical trap characterized
by an oscillator length $a_{r}=\sqrt{\hbar/m\omega}$, Eq. (\ref{TF0}) leads in
the TF limit to%

\begin{equation}
\frac{1}{2}m\omega^{2}r^{2}+gN\left\vert \psi(r)\right\vert ^{2}=\mu
,\hspace{0.6cm}0\leq r\leq R \label{TF1}%
\end{equation}
where the condensate radius $R=\sqrt{2\mu/\hbar\omega}\,a_{r}$ is determined
from the condition $\left\vert \psi(r)\right\vert ^{2}\geq0$, and the chemical
potential $\mu=\frac{1}{2}\left(  15Na/a_{r}\right)  ^{2/5}\hbar\omega$
follows from the normalization of $\psi(r)$.

In the opposite limit, when $N$ is small enough that the interaction energy
can be treated as a weak perturbation, one enters the\ (ideal gas)
perturbative regime. In this case, to the lowest order, $\psi(r)$ is given by
the harmonic oscillator ground state, $\psi(r)=(\pi a_{r}^{2})^{-3/4}%
\exp(-r^{2}/2a_{r}^{2})$, and the chemical potential satisfies%
\begin{equation}
(3/2)\hbar\omega+g\bar{n}=\mu\label{TF2}%
\end{equation}
where $\bar{n}=N/(\sqrt{2\pi}a_{r})^{3}$ is the mean atom density. Away from
these two limiting cases, in principle, one has to solve the GPE numerically.
Very few theoretical works have addressed the question of looking for
approximate analytical solutions valid in between the two analytically
solvable regimes. The most relevant proposals are based on a variational trial
wave function \cite{Fet1}, or on the semiclassical limit of the Wigner
phase-space distribution function of the condensate \cite{Vinas1}. However,
the practical usefulness of these approaches turns out to be somewhat limited
in comparison with the simple TF approximation.

In this work we address the above question from a different point of view. We
start from the usual TF approximation and modify it conveniently to account,
in a simple manner, for the zero-point energy contribution. This enables us to
derive simple and accurate extensions of the TF expressions that remain valid
for an arbitrary number of atoms in the mean-field regime. Specifically, we
obtain general analytical expressions for the ground-state properties of
spherical, cigar-shaped, and disk-shaped condensates that reduce to the
correct analytical formulas in both the TF and the perturbative regimes, and
remain valid and accurate in between these two limiting cases.

\begin{figure}
[ptb]
\begin{center}
\includegraphics[
height=5.3183cm,
width=7.9964cm
]%
{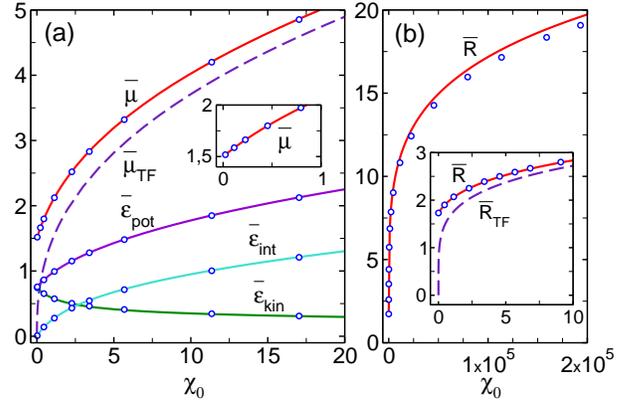}%
\caption{(Color online) Theoretical prediction for the ground-state properties
of spherical condensates (solid lines). The open circles are the exact
numerical results. For comparison purposes we have also included the TF
prediction (dashed lines).}%
\label{Fig1}%
\end{center}
\end{figure}

We begin by considering a BEC in a spherical trap. In principle, we start from
the TF relation of Eq. (\ref{TF1}). However, since we intend to apply this
equation to arbitrarily small condensates, we introduce a lower cutoff radius
$r_{0}$, defined\ through $\frac{1}{2}m\omega^{2}r_{0}^{2}=\frac{3}{2}%
\hbar\omega$, in order to be consistent with the fact that the contribution
from the harmonic oscillator energy cannot be smaller than the zero-point
energy. As for the small volume $V_{0}\sim a_{r}^{3}$ corresponding to $r\leq
r_{0}$, we do not aspire to get a precise knowledge of the wave function
therein. Instead, we content ourselves with an effective condensate density
$\bar{n}_{0}$ in that region. As we shall see, this is all that is needed to
obtain very accurate approximate formulas for most of the condensate
ground-state properties. Thus we start from the ansatz%

\begin{subequations}
\label{eq1ab}%
\begin{align}
\frac{1}{2}m\omega^{2}r^{2}+gN\left\vert \psi(r)\right\vert ^{2}  &
=\mu,\hspace{0.5cm}r_{0}<r\leq R\label{eq1a}\\
\frac{3}{2}\hbar\omega+g\sqrt{6/\pi}\,\bar{n}_{0}  &  =\mu,\hspace{0.5cm}0\leq
r\leq r_{0} \label{eq1b}%
\end{align}
with $\psi(r)=0$ for $r>R$. A renormalization constant $\kappa^{-1}\equiv
\sqrt{6/\pi}$ has been introduced in Eq. (\ref{eq1b}) to guarantee the correct
perturbative limit. In this limit $\mu\rightarrow\frac{3}{2}\hbar\omega$ and
$R\rightarrow r_{0}=\sqrt{3}\,a_{r}$. Under these circumstances, only Eq.
(\ref{eq1b}) contributes significantly to the chemical potential, and in this
case $\bar{n}_{0}=N/V_{0}$. This corresponds to a uniform spherical
condensate, defined in the finite volume $V_{0}$. In order for this uniform
density to produce the same chemical potential as the ground state of the
harmonic oscillator over the volume of the entire space it is only necessary
to renormalize the corresponding interaction strength by multiplying by
$\sqrt{6/\pi}$. Equations (\ref{eq1ab}) also yield the correct result in the
TF regime. This is mainly a consequence of the direct relation existing
between the number of particles and the size of a trapped BEC. For large
condensates, such that $\mu\gg\hbar\omega$, one has $R\gg r_{0}$ and, as a
result, the relative contribution from Eq. (\ref{eq1b}) to the normalization
integral that determines $\mu$ becomes negligible. Since we have renounced an
explicit expression for $\psi(r)$ in $V_{0}$, in this respect, our approach
cannot provide more information than the TF approach. Only when $R\gg r_{0}$
can we have a sufficiently precise knowledge of the wave function and, in this
case, it coincides with the TF wave function.

The chemical potential follows from the normalization of $\psi(r)$. After a
straightforward calculation one obtains%

\end{subequations}
\begin{equation}
\frac{1}{15}\overline{R}^{5}+\frac{\sqrt{3}}{2}\left(  \kappa-1\right)
\overline{R}^{2}-\frac{3\sqrt{3}}{2}\left(  \kappa-\frac{3}{5}\right)
=N\frac{a}{a_{r}},\label{eq2}%
\end{equation}
where $\overline{R}\equiv R/a_{r}$, and $\overline{\mu}=\frac{1}{2}%
\overline{R}^{2}$ is the chemical potential in units of $\hbar\omega$. As Eq.
(\ref{eq2}) shows, the ground-state properties depend on the sole parameter
$\chi_{0}\equiv Na/a_{r}$. When $\chi_{0}\gg1$ (TF limit) the above equation
leads to $\overline{\mu}=\frac{1}{2}\left(  15\chi_{0}\right)  ^{2/5}$, as
expected. The $\chi_{0}\ll1$ limit corresponds to the perturbative regime and,
in this case, one obtains $\overline{\mu}=3/2+\sqrt{2/\pi}\,\chi_{0}$, which
is nothing but the perturbative result (\ref{TF2}). For arbitrary $\chi_{0}$,
in principle one has to solve numerically the above quintic polynomial
equation (which has only one physically meaningful real solution). This is a
simple task that can be carried out with standard mathematical software
packages. We have found, however, a rather accurate approximate solution. It
can be shown that the expression%
\begin{equation}
\overline{R}^{2}=3+\left(  \frac{1}{\left(  15\chi_{0}\right)  ^{\frac{2}{5}%
}+\frac{5}{2}}+\frac{1}{\frac{7}{2}\chi_{0}^{11/15}+10}+\frac{\sqrt{\pi/2}%
}{2\chi_{0}}\right)  ^{-1}\label{eq2b}%
\end{equation}
satisfies Eq. (\ref{eq2}) with a residual error \cite{Error} smaller than
$0.7\%$ for any $\chi_{0}\in\lbrack0,\infty)$. Figures \ref{Fig1}(a) and
\ref{Fig1}(b) show, respectively, the predicted\ chemical potential,
$\overline{\mu}=\frac{1}{2}\overline{R}^{2}$, and condensate radius, obtained
from Eq. (\ref{eq2b}) (solid lines), along with the exact results obtained
from the numerical solution of the 3D GPE (open circles). For the numerical
calculation we have defined the radius through the condition $\left\vert
\psi(R)\right\vert ^{2}=0.05\left\vert \psi(0)\right\vert ^{2}$. With this
definition, Eq. (\ref{eq2b}) reproduces the numerical $R$ with a relative
error smaller than $3\%$ for any $\chi_{0}$. Most of the error, however, comes
from the region where $\chi_{0}\gg1$ (TF limit) because in that region
$R\rightarrow R_{\mathrm{TF}}$ and it rather satisfies $\left\vert
\psi(R)\right\vert ^{2}=0$. The accuracy with respect to the numerical
$\overline{\mu}$ is better than $0.5\%$.%

A straightforward calculation yields the mean-field interaction energy per
particle, $\overline{\epsilon}_{\mathrm{int}}\equiv\epsilon_{\mathrm{int}%
}/\hbar\omega\equiv E_{\mathrm{int}}/N\hbar\omega$,
\begin{align}
\overline{\epsilon}_{\mathrm{int}} &  =\frac{1}{8\chi_{0}}\left[  \frac
{8}{105}\overline{R}^{7}+\sqrt{3}\left(  \kappa-1\right)  \overline{R}%
^{4}\right.  \nonumber\\
&  \left.  -6\sqrt{3}\left(  \kappa-\frac{3}{5}\right)  \overline{R}%
^{2}+9\sqrt{3}\left(  \kappa-\frac{3}{7}\right)  \right]  .\label{eq3a}%
\end{align}
For $\chi_{0}\gg1$, one recovers the TF result, $\epsilon_{\mathrm{int}%
}=(2/7)\mu$. In the $\chi_{0}\ll1$ limit, using that $\overline{R}%
^{2}=3+2\sqrt{2/\pi}\,\chi_{0}-(1/\pi)(1/9+\sqrt{2/3\pi})\chi_{0}^{2}%
+O(\chi_{0}^{3})$ is a perturbative solution of Eq. (\ref{eq2}), one obtains
$\epsilon_{\mathrm{int}}=\chi_{0}\hbar\omega/\sqrt{2\pi}=g\bar{n}/2$, which
again is the correct result. Finally, the kinetic and potential energies can
be readily obtained in terms of the previous results by using the exact
relations \cite{RevStrin}
\begin{subequations}
\label{eq4ab}%
\begin{align}
\epsilon_{\mathrm{kin}} &  \equiv E_{\mathrm{kin}}/N=\mu
/2-(7/4)E_{\mathrm{int}}/N,\label{eq4a}\\
\epsilon_{\mathrm{pot}} &  \equiv E_{\mathrm{pot}}/N=\mu
/2-(1/4)E_{\mathrm{int}}/N.\label{eq4b}%
\end{align}
In Fig. \ref{Fig1}(a) we show the theoretical prediction for $\overline
{\epsilon}_{\mathrm{int}}$, $\overline{\epsilon}_{\mathrm{kin}}$, and
$\overline{\epsilon}_{\mathrm{pot}}$, obtained from Eqs. (\ref{eq2b}%
)--(\ref{eq4ab}) (solid lines), along with the exact numerical results (open circles).

Next we consider a BEC confined in a cigar-shaped magnetic trap with
oscillator lengths $a_{\bot}=\sqrt{\hbar/m\omega_{\bot}}$ and $a_{z}%
=\sqrt{\hbar/m\omega_{z}}$ and an aspect ratio $\lambda=\omega_{z}%
/\omega_{\bot}\ll2$. We shall restrict ourselves to the mean-field regime,
which requires $N\lambda a_{\bot}^{2}/a^{2}\gg1$ \cite{Petrov1,Dunj1,Strin1}.
As before, we start from the usual TF expression, which we assume to be valid
up to a minimum radial distance $r_{\bot}^{0}=\sqrt{2}\,a_{\bot}$, determined
from the condition that the contribution from the radial harmonic oscillator
energy should not be smaller than $\hbar\omega_{\bot}$. This defines an outer
region $V_{+}\equiv\left\{  (r_{\bot},z)\colon\,r_{\bot}^{2}/R^{2}%
+z^{2}/Z_{\mathrm{TF}}^{2}\leq1\;\wedge\;r_{\bot}>r_{\bot}^{0}\right\}  $,
which is nothing but the usual TF ellipsoidal density cloud, truncated at
$r_{\bot}=r_{\bot}^{0}$. Note that unlike what happens with the condensate
radius $R=\sqrt{2\mu/\hbar\omega_{\bot}}\,a_{\bot}$, which remains the same,
now the axial condensate half-length $Z=\sqrt{2(\mu/\hbar\omega_{\bot}%
-1)}\,a_{z}/\sqrt{\lambda}$ coincides with the TF value $Z_{\mathrm{TF}}%
$\ only in the limit $\mu/\hbar\omega_{\bot}\gg1$. For large condensates, when
$\mu\gg\hbar\omega_{\bot}$ (TF regime), this is the only region that
contributes significantly. On the contrary, in the perturbative regime, as
$\mu\rightarrow\hbar\omega_{\bot}$ most of the contribution comes from the
inner cylinder $V_{-}\equiv\left\{  (r_{\bot},z)\colon\,r_{\bot}\leq r_{\bot
}^{0}\;\wedge\;|z|\leq Z\right\}  $. In this case, if $a\ll a_{\bot}$, the
transverse dynamics becomes frozen in the radial ground state of the harmonic
trap and the condensate wave function can be factorized as $\psi(r_{\bot
},z)=\varphi(r_{\bot})\phi(z)$, with $\varphi(r_{\bot})=(\pi a_{\bot}%
^{2})^{-1/2}\exp(-r_{\bot}^{2}/2a_{\bot}^{2})$. This corresponds to a
mean-field quasi-1D condensate. Substituting then in Eq. (\ref{TF0}) and
integrating out the radial dynamics, one finds
\end{subequations}
\begin{equation}
\hbar\omega_{\bot}+\frac{1}{2}m\omega_{z}^{2}z^{2}+g_{\mathrm{1D}}%
N|\phi(z)|^{2}=\mu,\label{eq5}%
\end{equation}
where $g_{\mathrm{1D}}=g/2\pi a_{\bot}^{2}$ \cite{Olsha2}, and we have used
that $\mu\sim\hbar\omega_{\bot}\gg\frac{1}{2}\hbar\omega_{z}$ to neglect the
axial kinetic energy. Note that $g_{\mathrm{1D}}$ can be conveniently
rewritten as $g\bar{n}_{2}$ with $\bar{n}_{2}=1/\pi(r_{\bot}^{0})^{2}$,
indicating that one can account for the contribution from the radial ground
state by using a uniform mean density per unit area normalized to unity in
$V_{-}$. Guided by these simple ideas, we then propose the following ansatz:%

\begin{align*}
\frac{1}{2}m\omega_{\bot}^{2}r_{\bot}^{2}+\frac{1}{2}m\omega_{z}^{2}%
z^{2}+gN\left\vert \psi(r_{\bot},z)\right\vert ^{2}  &  =\mu,\hspace
{0.5cm}\mathbf{r}\in V_{+}\\
\hbar\omega_{\bot}+\frac{1}{2}m\omega_{z}^{2}z^{2}+gN\bar{n}_{2}|\phi(z)|^{2}
&  =\mu,\hspace{0.5cm}\mathbf{r}\in V_{-}%
\end{align*}
with $\psi=0$ elsewhere. The normalization of $\psi$ leads to%

\begin{equation}
\frac{1}{15}(\sqrt{\lambda}\,\overline{Z})^{5}+\frac{1}{3}(\sqrt{\lambda
}\,\overline{Z})^{3}=N\lambda\frac{a}{a_{\bot}}, \label{eq7}%
\end{equation}
where $\overline{Z}\equiv Z/a_{z}$ and $\overline{R}\equiv R/a_{\bot}$. The
chemical potential $\overline{\mu}\equiv\mu/\hbar\omega_{\bot}$ is given by
$\overline{\mu}=1+\frac{1}{2}(\sqrt{\lambda}\,\overline{Z})^{2}$. Now the
relevant parameter determining the ground-state properties is $\chi_{1}\equiv
N\lambda a/a_{\bot}$. When $\chi_{1}\gg1$ (TF regime), Eq. (\ref{eq7}) leads
to $\overline{\mu}=\frac{1}{2}(15\chi_{1})^{2/5}$ and $\overline{Z}%
=\lambda^{-1/2}(15\chi_{1})^{1/5}$. When $\chi_{1}\ll1$ (mean-field quasi-1D
regime), one obtains $\overline{\mu}=1+\frac{1}{2}(3\chi_{1})^{2/3}$ and
$\overline{Z}=\lambda^{-1/2}(3\chi_{1})^{1/3}$, in agreement with previous
results \cite{Dunj1,Strin1}. In general, for arbitrary $\chi_{1}$, an
approximate solution satisfying Eq. (\ref{eq7}) with a residual error less
than $0.75\%$ for any $\chi_{1}\in\lbrack0,\infty)$ is given \ by%

\begin{equation}
\sqrt{\lambda}\,\overline{Z}=\left(  \frac{1}{\left(  15\chi_{1}\right)
^{\frac{4}{5}}+\frac{1}{3}}+\frac{1}{57\chi_{1}+345}+\frac{1}{(3\chi
_{1})^{\frac{4}{3}}}\right)  ^{-\frac{1}{4}}\label{eq8}%
\end{equation}
The mean-field interaction energy $\overline{\epsilon}_{\mathrm{int}}%
\equiv\epsilon_{\mathrm{int}}/\hbar\omega_{\bot}$ is%

\begin{equation}
\overline{\epsilon}_{\mathrm{int}}=\frac{1}{15\chi_{1}}\left(  (\sqrt{\lambda
}\,\overline{Z})^{5}+\frac{1}{7}(\sqrt{\lambda}\,\overline{Z})^{7}\right)
.\label{eq9}%
\end{equation}
For $\chi_{1}\gg1$, Eq. (\ref{eq9})\ reduces to $\epsilon_{\mathrm{int}%
}=(2/7)\mu$, while for $\chi_{1}\ll1$, it leads to $\epsilon_{\mathrm{int}%
}=(2/5)(\mu-\hbar\omega_{\bot})$, which again are the correct analytical
limits. As for the condensate density per unit length, $n_{1}(z)\equiv
N\int2\pi r_{\bot}dr_{\bot}\left\vert \psi(r_{\bot},z)\right\vert ^{2}$, after
a straightforward calculation one finds
\begin{equation}
n_{1}(z)=\frac{(\sqrt{\lambda}\;\overline{Z})^{2}}{4a}\left(  1-\frac{z^{2}%
}{Z^{2}}\right)  +\frac{(\sqrt{\lambda}\;\overline{Z})^{4}}{16a}\left(
1-\frac{z^{2}}{Z^{2}}\right)  ^{2}\label{eq10}%
\end{equation}
The first term is the contribution from $V_{-}$ and thus it is the only one
that contributes significantly in the $\chi_{1}\ll1$ limit. On the contrary,
the second term, which is the contribution from $V_{+}$, gives the dominant
contribution in the $\chi_{1}\gg1$ limit, in good agreement with previous
results \cite{Strin1}.%

\begin{figure}
[ptb]
\begin{center}
\includegraphics[
height=5.3183cm,
width=7.9964cm
]%
{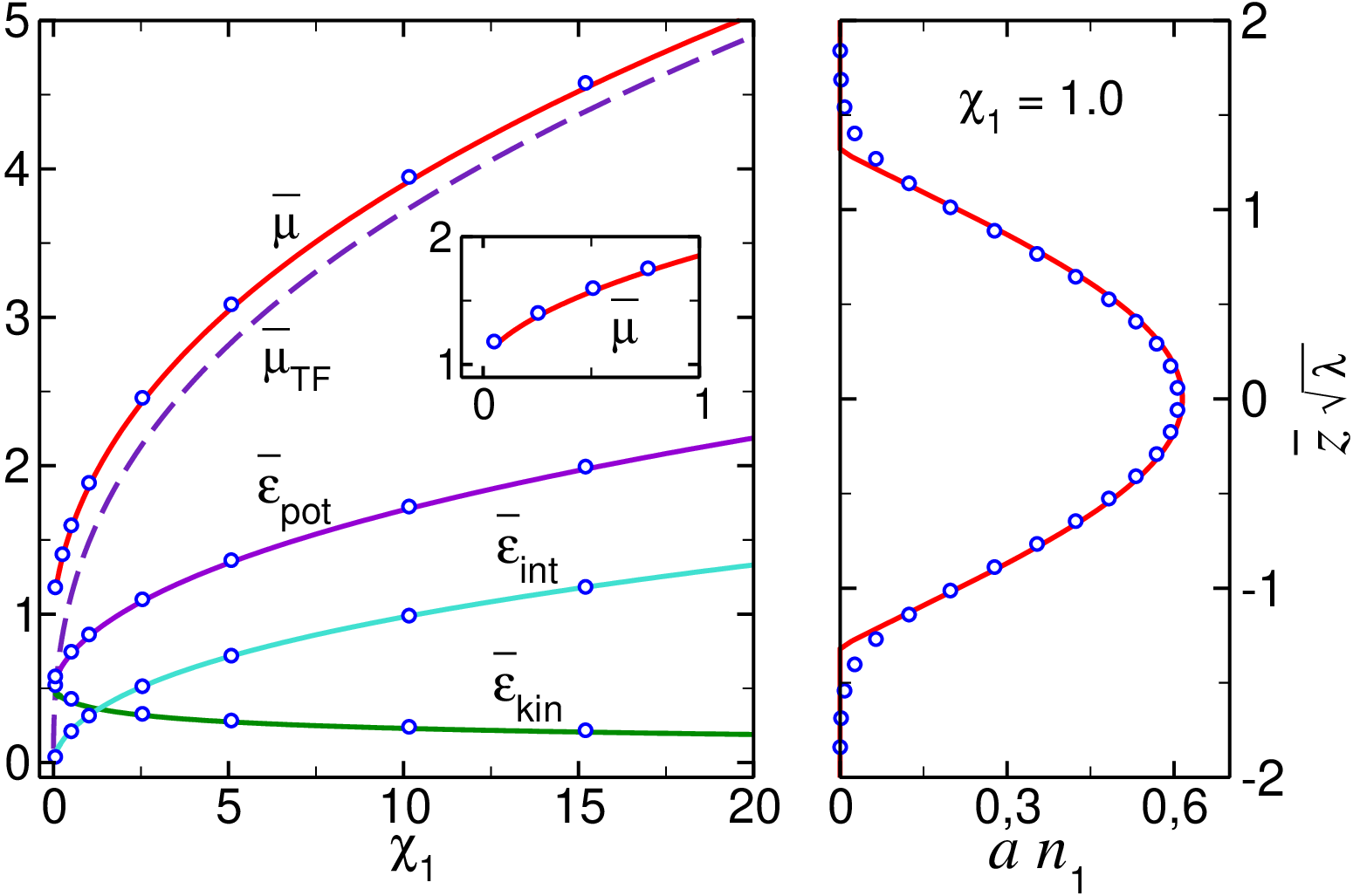}%
\caption{(Color online) Theoretical prediction for the ground-state properties
of arbitrary cigar-shaped condensates with $\lambda\ll2$ (solid lines). The
open circles are exact numerical results obtained with $\lambda=0.2$. The
dashed line is the TF prediction.}%
\label{Fig2}%
\end{center}
\end{figure}

Figure \ref{Fig2} shows the theoretical predictions for the ground-state
properties of arbitrary cigar-shaped condensates with $\lambda\ll2$, obtained
from Eqs. (\ref{eq4ab}) and (\ref{eq8})--(\ref{eq10}) (solid lines), along
with exact numerical results (open circles).

Finally, we consider a BEC in a disk-shaped trap with $\lambda\gg2$ and
$a_{z}\gg a$. In this case, in the mean-field peturbative regime, which occurs
when $\chi_{2}\equiv Na/\lambda^{2}a_{z}\ll1$, the system reduces to a
quasi-2D condensate satisfying%

\begin{equation}
\frac{1}{2}\hbar\omega_{z}+\frac{1}{2}m\omega_{\bot}^{2}r_{\bot}%
^{2}+g_{\mathrm{2D}}N|\varphi(r_{\bot})|^{2}=\mu,\label{eq11}%
\end{equation}
with $g_{\mathrm{2D}}=g/\sqrt{2\pi}\,a_{z}$ \cite{Petrov2}. We then rewrite
$g_{\mathrm{2D}}$ as $g\kappa_{2}^{-1}\bar{n}_{1}$, where $\bar{n}%
_{1}=1/2a_{z}$ is a uniform mean density per unit length and $\kappa_{2}%
^{-1}\equiv\sqrt{2/\pi}$ is the appropriate renormalization factor, and
propose the following ansatz:%

\begin{align}
\frac{1}{2}m\omega_{z}^{2}z^{2}+\frac{1}{2}m\omega_{\bot}^{2}r_{\bot}%
^{2}+gN\left\vert \psi(r_{\bot},z)\right\vert ^{2} &  =\mu,\hspace
{0.5cm}\mathbf{r}\in V_{+}\nonumber\\
\frac{1}{2}\hbar\omega_{z}+\frac{1}{2}m\omega_{\bot}^{2}r_{\bot}^{2}%
+g\kappa_{2}^{-1}N\bar{n}_{1}|\varphi(r_{\bot})|^{2} &  =\mu,\hspace
{0.5cm}\mathbf{r}\in V_{-}\label{eq11b}%
\end{align}
with $\psi=0$ elsewhere. In the above equations, $V_{+}\equiv\left\{
(r_{\bot},z)\colon\,r_{\bot}^{2}/R_{\mathrm{TF}}^{2}+z^{2}/Z^{2}\leq
1\,\wedge\,|z|>z_{0}\right\}  $ and $V_{-}\equiv\left\{  (r_{\bot}%
,z)\colon\,r_{\bot}\leq R\;\wedge\;|z|\leq z_{0}\right\}  $, where
$z_{0}=a_{z}$, $R_{\mathrm{TF}}=\sqrt{2\mu/\hbar\omega_{z}}\sqrt{\lambda
}\,a_{\bot}$, $R=\sqrt{2(\mu/\hbar\omega_{z}-1/2)}\sqrt{\lambda}\,a_{\bot}$,
and $Z=\sqrt{2\mu/\hbar\omega_{z}}\,a_{z}$. More precisely, one expects
$\kappa_{2}^{-1}\rightarrow\sqrt{2/\pi}$ in the perturbative regime ($\chi
_{2}\ll1$), while $\kappa_{2}^{-1}\rightarrow1$ in the TF regime ($\chi_{2}%
\gg1$). The final results are not very sensitive to the specific functional
form of $\kappa_{2}^{-1}$. We thus propose one of the simplest possibilities:%

\begin{align}
\kappa_{2}^{-1}(\chi_{2}) &  \equiv\sqrt{2/\pi}+\Theta(\chi_{2}%
-0.1)\nonumber\\
\times &  \left(  1-\sqrt{2/\pi}\right)  \left(  1-\frac{R_{\mathrm{TF}}%
(\chi_{2}=0.1)}{R_{\mathrm{TF}}(\chi_{2})}\right)  ,\label{eq12}%
\end{align}
where $\Theta(x)$ is the Heaviside function and $R_{\mathrm{TF}}(\chi
_{2})=(15\chi_{2})^{1/5}a_{\bot}$ is the TF radius. The normalization of
$\psi$ yields%

\begin{equation}
\frac{1}{15}\overline{Z}^{5}+\frac{1}{8}(\kappa_{2}-1)\frac{\overline{R}^{4}%
}{\lambda^{2}}-\frac{\overline{R}^{2}}{6\lambda}-\frac{1}{15}=\frac
{Na}{\lambda^{2}a_{z}}, \label{eq13}%
\end{equation}
where $\overline{Z}\equiv Z/a_{z}$, $\overline{R}\equiv R/a_{\bot}$, and
$\overline{Z}^{2}-\overline{R}^{2}/\lambda=1$. The chemical potential is
$\overline{\mu}\equiv\mu/\hbar\omega_{z}=\frac{1}{2}(1+\overline{R}%
^{2}/\lambda)$.

For $\chi_{2}\gg1$ Eq. (\ref{eq13}) leads to the usual TF results, while for
$\chi_{2}\ll1$ (mean-field quasi-2D regime), one obtains $\overline{\mu
}=1/2+(2\sqrt{2/\pi}\chi_{2})^{1/2}$ and $\overline{R}=\lambda^{1/2}%
(8\sqrt{2/\pi}\chi_{2})^{1/4}$. An approximate solution that satisfies Eq.
(\ref{eq13}) with a residual error less than $0.95\%$ for any $\chi_{2}%
\in\lbrack0,\infty)$ is given \ by%
\begin{equation}
\overline{R}_{\lambda}\equiv\overline{R}/\sqrt{\lambda}=\left[  \left(
1/15\chi_{2}\right)  ^{8/5}+\left(  \kappa_{2}/8\chi_{2}\right)  ^{2}\right]
^{-1/8} \label{eq14}%
\end{equation}

After some calculation one finds the following expressions for the mean-field
interaction energy $\overline{\epsilon}_{\mathrm{int}}\equiv\epsilon
_{\mathrm{int}}/\hbar\omega_{z}$ and the condensate density per unit area
$n_{2}(r_{\bot})$:%

\begin{equation}
\overline{\epsilon}_{\mathrm{int}}=\frac{1}{8\chi_{2}}\left(  \frac
{8\overline{Z}^{7}}{105}+\xi\frac{\overline{R}_{\lambda}^{6}}{6}%
-\frac{\overline{R}_{\lambda}^{4}}{3}-\frac{4\overline{R}_{\lambda}^{2}}%
{15}-\frac{8}{105}\right)  \label{eq15}%
\end{equation}

\begin{equation}
n_{2}(r_{\bot})=\frac{\xi\left[  2\overline{\mu}_{z}(r_{\bot})-1\right]
}{4\pi aa_{z}}+\frac{\left[  2\overline{\mu}_{z}(r_{\bot})\right]  ^{3/2}%
-1}{6\pi aa_{z}}, \label{eq16}%
\end{equation}
where $\xi\equiv(\kappa_{2}-1)$ and $2\overline{\mu}_{z}(r_{\bot}%
)\equiv1+\overline{R}_{\lambda}^{2}\left(  1-r_{\bot}^{2}/R^{2}\right)  $.%

\begin{figure}
[ptb]
\begin{center}
\includegraphics[
height=5.3183cm,
width=7.9964cm
]%
{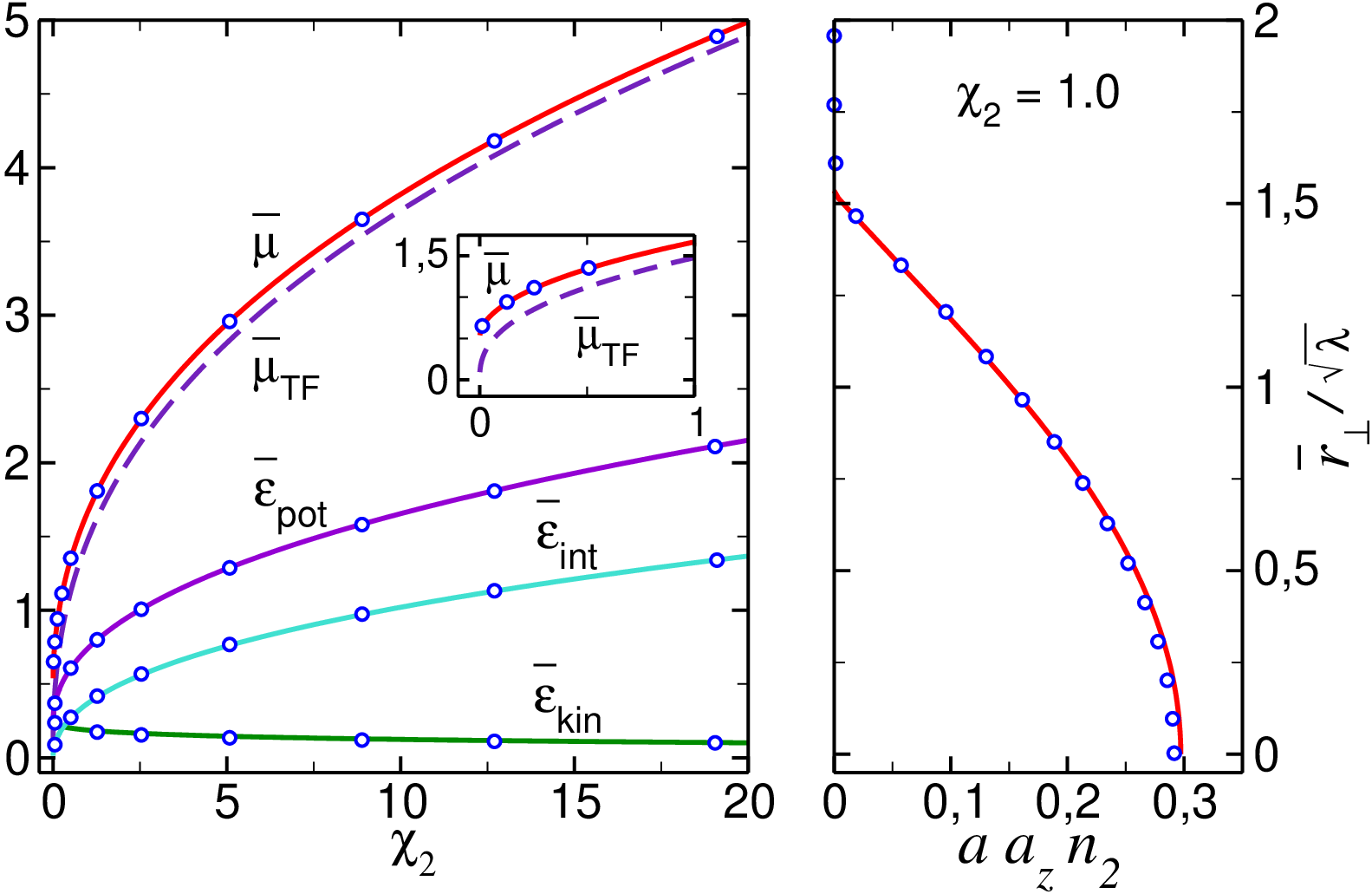}%
\caption{(Color online) Theoretical prediction for the ground-state properties
of arbitrary disk-shaped condensates with $\lambda\gg2$ (solid lines). The
open circles are exact numerical results obtained with $\lambda=20$. The
dashed line is the TF prediction.}%
\label{Fig3}%
\end{center}
\end{figure}

In Fig. \ref{Fig3} we show the ground-state properties of arbitrary
disk-shaped condensates with $\lambda\gg2$, obtained from our analytical
formulas [Eqs. (\ref{eq4ab}) and (\ref{eq14})--(\ref{eq16})] (solid lines),
along with exact numerical results (open circles).

In conclusion, modifying the usual TF approximation conveniently to account
for the zero-point energy contribution, we have derived general analytical
expressions for the ground-state properties of spherical, cigar-shaped, and
disk-shaped condensates that reduce to the correct analytical formulas in both
the TF and the mean-field perturbative regimes and remain valid and accurate
in between these two limiting cases. Mean-field quasi-1D and -2D condensates
appear as simple particular cases of our formulation.

\begin{acknowledgments}
This work has been supported by MEC (Spain) and FEDER fund (EU) (Contract No. Fis2005-02886).
\end{acknowledgments}

\bigskip


\begin{thebibliography}{99}                                                                                               %


\bibitem {BEC1}M. H. Anderson \emph{et al.}, Science \textbf{269}, 198 (1995).

\bibitem {BEC2}K. B. Davis \emph{et al.}, Phys. Rev. Lett. \textbf{75}, 3969 (1995).

\bibitem {BEC3}C. C. Bradley \emph{et al.}, Phys. Rev. Lett. \textbf{78}, 985 (1997).

\bibitem {Baym1}G. Baym and C. J. Pethick, Phys. Rev. Lett. \textbf{76}, 6 (1996).

\bibitem {RevStrin}For a review see, for example, F. Dalfovo, S. Giorgini, L.
P. Pitaevskii, and S. Stringari, Rev. Mod. Phys. \textbf{71}, 463 (1999).

\bibitem {Fet1}A. L. Fetter, J. Low Temp. Phys. \textbf{106}, 643 (1997).

\bibitem {Vinas1}P. Schuck and X. Vi\~{n}as, Phys. Rev. A \textbf{61}, 43603 (2000).

\bibitem {Error}Given $P(R)=\chi$, we define the residual error associated
with the approximate solution $R_{\varepsilon}$ as $[P(R_{\varepsilon}%
)-\chi]/\chi$.

\bibitem {Petrov1}D. S. Petrov \emph{et al.}, Phys. Rev. Lett. \textbf{85},
3745 (2000).

\bibitem {Dunj1}V. Dunjko, V. Lorent, and M. Olshanii, Phys. Rev. Lett.
\textbf{86}, 5413 (2001).

\bibitem {Strin1}C. Menotti and S. Stringari, Phys. Rev. A \textbf{66}, 043610 (2002).

\bibitem {Olsha2}M. Olshanii, Phys. Rev. Lett. \textbf{81}, 938 (1998).

\bibitem {Petrov2}D. S. Petrov \emph{et al.}, Phys. Rev. Lett. \textbf{84},
2551 (2000).
\end{thebibliography}
\end{document}